\documentclass{article}

\usepackage{amsmath}

\usepackage{xcolor}
\usepackage[numbers]{natbib}
\usepackage[english]{babel}

\usepackage{hyperref}         
\hypersetup{                    
  colorlinks,
  linkcolor={green!80!black},
  citecolor={red!70!black},
  urlcolor={blue!70!black}
}
\usepackage{cleveref}
\usepackage{float}
\usepackage{listings}
\usepackage{subcaption}
\usepackage{csquotes}

\newfloat{lstfloat}{htbp}{lop}
\floatname{lstfloat}{Listing}
\DeclareCaptionSubType*[arabic]{lstfloat}
\crefname{lstfloat}{listing}{listings}
\Crefname{lstfloat}{Listing}{Listings}
\crefname{sublstfloat}{listing}{listings}
\Crefname{sublstfloat}{Listing}{Listings}

\newcommand{\ie}{\textit{i}.\textit{e}.}
\newcommand{\eg}{\textit{e}.\textit{g}.}
\newcommand{\etal}{\textit{et} \textit{al}.}

\begin{document}


\title{ROMEO: A Binary Vulnerability Detection Dataset for Exploring Juliet through the Lens of Assembly Language}
\author{Clemens-Alexander Brust, Tim Sonnekalb, Bernd Gruner\\DLR Institute of Data Science\\Jena, Germany\\%
\texttt{firstname.lastname@dlr.de}}
\date{}

\maketitle{}

\section*{Abstract}
\paragraph{Context}
Automatic vulnerability detection on C/C++ source code has benefitted from the introduction of machine learning to the field, with many recent publications targeting this combination. In contrast, assembly language or machine code artifacts receive less attention, although there are compelling reasons to study them. They are more representative of what is executed, more easily incorporated in dynamic analysis, and in the case of closed-source code, there is no alternative.

\paragraph{Objective}
We evaluate the representative capability of assembly language compared to C/C++ source code for vulnerability detection. Furthermore, we investigate the role of call graph context in detecting function-spanning vulnerabilities. Finally, we verify whether compiling a benchmark dataset compromises an experiment's soundness by inadvertently leaking label information.

\paragraph{Method}
We propose ROMEO, a publicly available, reproducible and reusable binary vulnerability detection benchmark dataset derived from the synthetic Juliet test suite. Alongside, we introduce a simple text-based assembly language representation that includes context for function-spanning vulnerability detection and semantics to detect high-level vulnerabilities. It is constructed by disassembling the \texttt{.text} segment of the respective binaries.

\paragraph{Results}
We evaluate an x86 assembly language representation of the compiled dataset, combined with an off-the-shelf classifier. It compares favorably to state-of-the-art methods, including those operating on the full C/C++ code. Including context information using the call graph improves detection of function-spanning vulnerabilities. There is no label information leaked during the compilation process.

\paragraph{Conclusion}
Performing vulnerability detection on a compiled program instead of the source code is a worthwhile tradeoff. While certain information is lost, e.g., comments and certain identifiers, other valuable information is gained, e.g., about compiler optimizations.

\section{Introduction}
Machine learning has advanced automatic vulnerability detection significantly compared to traditional static analysis security testing (SAST) tools \cite{Li2018VulDeePeckerDeepLearning}.
The focus of recent methods is mainly on C/C++ source code, where many vulnerabilities can be detected reliably, but others cannot \cite{Chakraborty2021DeepLearningBased}.
We argue that, at least for C/C++ software projects, but also other languages, the compiler's output warrants the same attention as the source code, if not more.
The machine code is more closely related to the actual state transitions during execution than the original source code, and SAST tools often fail to consider implementation details of compilers \cite{Balakrishnan2010WYSINWYXWhatYou}.
Moreover, source code is not always available for third party software, which nevertheless has to be audited.
Finally, a method that works directly on assembly language or machine code can easily be transferred to a dynamic environment, \eg{} for runtime monitoring of JIT compilers, and it supports all compiled higher-level languages by design.

While there is recent work in the area of vulnerability detection on machine code and assembly language (see \cref{sec:relwvulndetasm}), there is no reproducible, publicly available dataset for this task.
Previous work suffers from further limitations (see \cref{sec:relatedwork}).
To the best of our knowledge, there is also no binary vulnerability method that incorporates the semantics and context information necessary to detect more abstract vulnerabilities spanning multiple functions.
To advance the state of the art into this direction, we contribute the following:
\begin{description}
  \item[The ROMEO dataset:] a binary vulnerability detection benchmark dataset based on the Juliet test suite \cite{JulietUserGuide} version 1.3, with approx. 168k examples labeled as one of 91 Common Weakness Enumeration (CWE, \cite{mitrecwe}) categories. We publish the source code of the entire processing pipeline, which is reproducible and configurable to meet the needs of other researchers.
  \item[The ROMEO representation:] a simple text representation of disassembled binaries suitable for various sequence classifiers. It incorporates context for across-function vulnerabilities and preserves semantics to identify API calls, while still preventing label leakage.
  \item[Experiments:] using an off-the-shelf Transformer approach \cite{Feng2020CodeBERTPretrained}, we show the efficacy of our assembly language representation, which compares favorably even to state-of-the-art methods that have access to the full source code, and highlight the relative strengths and weaknesses in qualitative analyses.
\end{description}

The remainder of this work is structured as follows: we first present related work in \cref{sec:relatedwork}.
We then give a detailed description of the dataset creation process and our representation in \cref{sec:extractingassembly}.
Research questions are posed and answered experimentally in \cref{sec:experiments}.
We finally offer a brief conclusion and an outlook towards future research in \cref{sec:conclusion}.

\section{Related Work}\label{sec:relatedwork}
In this section, we review previous work related to our investigation.
We focus on two main areas. First, we consider approaches to vulnerability detection that use an assembly language or machine code representation. Second, we explore work that uses the Juliet test suite as a benchmark or evaluation dataset.

\subsection{Vulnerability Detection using Assembly Language Representations}\label{sec:relwvulndetasm}

Lee \etal{} \cite{Lee2017LearningBinaryCode} combine a bespoke encoding of assembly language instructions called Instruction2vec~\cite{Lee2019Instruction2vecEfficientPreprocessor} with a deep learning model \enquote{Text-CNN} to detect vulnerabilities in binary code.
The encoding represents each instruction as a vector of a fixed length.
Individual components of the instruction, \ie{} the opcode and operand fragments, are encoded using a custom word2vec~\cite{Mikolov2013EfficientEstimationWord} model.
Their model detects CWE-121 (Stack Overflow) vulnerabilities in Juliet with an accuracy of 96.1\% compared to an off-the-shelf word2vec at 94.2\%. We replicate their setup in our evaluation for comparison.

BVDetector~\cite{Tian2020BVDetectorProgramSlice} operates based on pre-extracted program slices.
It relies on a per-token word2vec encoding.
While the authors test a variety of neural networks to classify the encoded slices, they find that a BGRU~\cite{Cho2014LearningPhraseRepresentations} performs best.
They measure the performance of their method on program slices extracted from a subset of the Juliet test suite concerning memory corruption and number handling vulnerabilities and report an accuracy of 96.7\%.
We construct a subset using the same criteria to compare our approach to BVDetector as their dataset is not publicly available.

The notion of program slices~\cite{Li2018VulDeePeckerDeepLearning,Li2021SySeVRFrameworkUsing} is extended to assembly language by Li \etal{}~\cite{Li2021VulnerabilityDetectionSystem}.
They also incorporate a combined representation of source code and assembly called \enquote{hybrid slices}.
The hybrid slice method is evaluated on an aggregated subset of Juliet, where it reaches 96.9\% accuracy compared to BVDetector at 88.9\%.

Le \etal{} propose a Maximal Divergence Sequential Auto-Encoder~\cite{Le2019MaximalDivergenceSequential} and rely on a fixed encoding of opcodes and a histogram-like encoding of operands to represent assembly language instructions. The autoencoder is evaluated on the dataset introduced alongside VulDeePecker~\cite{Li2018VulDeePeckerDeepLearning}, and outperforms it slightly with an accuracy of 85.3\% vs. 83.5\%.

In \cite{Afanador2021BenchmarkFrameworkSupport,Afanador2020RepresentativenessBenchmarkVulnerability}, a dataset called BVATT is constructed similarly to ours. It is intended as a reliable benchmark for binary vulnerability detection methods. However, it is no longer publicly available.

All the aforementioned methods choose different learned or hand-crafted encodings of the machine code.
However, none attempt to leverage the human-readable mnemonic representation of assembly as we propose in this work.

\subsection{Juliet Test Suite as a Benchmark for Vulnerability Detection}
In the following, we provide a brief overview of related work where the Juliet test suite is used as a benchmark dataset outside its intended SAST application.

Russel \etal{}~\cite{Russell2018AutomatedVulnerabilityDetection} augment Juliet with examples mined from GitHub repositories and Debian packages. The mined examples are annotated using the static analysis tools Clang, Cppcheck and Flawfinder.
While the tool selection is diverse, it cannot provide a replacement for manually labeled data. Machine learning-based methods can simply learn the rules embedded in these tools, and thus will produce similar errors. Juliet, while synthetic, is almost by definition always labeled correctly.

Li \etal{} evaluate their method VulDeePecker~\cite{Li2018VulDeePeckerDeepLearning} on a combination of CWEs 119 and 399 of the Juliet test suite and vulnerabilities in open source projects listed in the National Vulnerability Database (NVD).
This dataset is compiled and used by Le \etal{} in~\cite{Le2019MaximalDivergenceSequential} to validate their binary vulnerability detector (see above).
The NVD provides compelling real-world examples, but introduces label quality concerns into the dataset due to the complex mining process involved.
Furthermore, the dataset is not suitable for our evaluation because its individual examples are not compilable on their own.
Li \etal{} refine and extend the combination of Juliet and examples from the NVD to evaluate SySeVR~\cite{Li2021SySeVRFrameworkUsing}.

BVDetector~\cite{Tian2020BVDetectorProgramSlice} uses a subset of the Juliet test suite without any additions (see \cref{sec:relwvulndetasm} for further details).
A similar approach, but combined with the C/C++ source code, is taken by Li \etal{}~\cite{Li2021VulnerabilityDetectionSystem}.
Conversely, our evaluation uses a much larger subset, missing only Windows-specific weaknesses.
The Juliet test suite is further used as an alternative training and evaluation dataset in the \textsc{ReVeal} study~\cite{Chakraborty2021DeepLearningBased} by Chakraborty \etal{}, who criticize the high number of duplicates compared to real-world datasets.
We describe our approach to this matter in \cref{sec:duplicateelimination}.

\section{Extracting an Assembly Language Dataset from the Juliet Test Suite}\label{sec:extractingassembly}
This section details the steps required to extract an assembly language dataset from the Juliet test suite that is suitable for processing by a machine learning model.
The resulting dataset should fulfill the following requirements, which will serve as our design goals throughout the process:
\begin{description}
  \item[Coverage] The dataset should cover as large a fraction of the Juliet test suite as possible. While previous work \cite{Lee2017LearningBinaryCode,Tian2020BVDetectorProgramSlice} focuses on subsets of the test suite to manage complexity, or uses static analysis rules to extract examples \cite{Li2018VulDeePeckerDeepLearning}, this work covers the largest possible fraction of the entire test suite that does not compromise the following two properties and is compatible with Linux environments.
  \item[Context] While our goal is to provide examples on a function level, a single function can be vulnerable or not depending on the content of related functions or data. The dataset should provide appropriate context in terms of relevant functions for each example.
  \item[No Label Leakage] Because of the very methodical construction of the test cases in Juliet, there are various ways in which label information can leak into examples. Symbols include label information in their names, \eg{}, when they end in \texttt{good} or \texttt{bad}. Context can also leak information, \eg{}, the presence of any context is enough to distinguish a primary good function from a primary bad function in certain flow variants. Label leakage can lead to overestimation of a machine learning model's predictive performance during validation as the model exploits inadvertent correlations between examples and labels. Consequently, a model trained on a \enquote{leaky} dataset will exhibit worse performance in real-world applications, where these correlations do not exist. Hence, no label leakage should occur.
\end{description}

\subsection{Preparations}
Before we can extract examples, the Juliet test suite requires some preparations, which we discuss in the following.

\subsubsection{Modifications}
As a sanity check, the support file \texttt{io.c}, which we need to include for context, contains 18 empty functions by the names of \texttt{good1}-\texttt{good9} and \texttt{bad1}-\texttt{bad9}.
We have to remove these functions from the support file because they would be present in every testcase after linking, and could be confused with actual examples.
Ignoring empty functions entirely is not feasible, as there are testcases where empty functions are present on purpose, \eg{}, concerning CWE-570 (Expression Always False).
Hence, this modification is unavoidable. Aside from these support functions, we do not modify the Juliet source code in any way.

\subsubsection{Compilation and Linking}
We compile all testcases individually with GCC 11.2.0 using the provided Makefile generation script.
Since our platform of choice is Linux, the script does not create Makefiles for Windows-specific CWEs such as CWE-247 (Reliance on DNS Lookups in a Security Decision).
Compiling testcases individually results in single object files per translation unit, which we then link with the compiled support unit \texttt{io.c} because functions such as \texttt{printLine} are used in the testcases and could provide relevant context.
Testcases consisting of multiple translation units are linked into one object file, together with the support unit.

Overall, this process results in object files for 41,812 testcases covering 91 CWEs, with the remainder consisting of Windows-specific testcases.

\subsection{From C/C++ to Assembly Language}
To obtain an assembly language representation of C/C++ sources, there are two obvious options.
First, one can compile the sources directly into assembly language, skipping the assembly into machine code.
Second, one can compile the sources fully into machine code, link them, and then disassemble the resulting binaries.
We choose the latter approach, as it can in principle reveal more information about optimizations, \eg{}, link-time optimization (LTO), as well as specific instruction encodings.
It also more closely resembles a black-box analysis situation where source code is not available.

We disassemble each testcase's linked object file using \texttt{objdump} configured to produce Intel syntax assembly.
C++ symbols are demangled.
Moreover, we extract static and dynamic symbol tables for each testcase to later distinguish global and local functions.
The disassembly is parsed including addresses, binary representations, sections etc. and compared to the output of \texttt{capstone} on a per-instruction basis to ensure a correct disassembly.

\subsection{Extracting Examples}\label{sec:extractingexamples}
In the following, we construct the ROMEO assembly language representation from the disassembly in a number of steps, including modifying operands and symbols, and selecting relevant context.
Furthermore, we describe the mapping from testcases in Juliet to examples in the ROMEO dataset.

\subsubsection{Symbol Representation}\label{sec:symbolrepresentation}
We build a scrambling table for each testcase, mapping each local symbol to a unique random name in the pattern of \texttt{lc000}-\texttt{lc999} to prevent label leakage, \eg{}, from function names ending in \texttt{good} or \texttt{bad}.
Global symbols are not renamed, as they contain no label information, but are crucial for more abstract vulnerability detection, \eg{}, \texttt{memcpy}.

Many instructions contain addresses in the operands, which \texttt{objdump} annotates with corresponding symbols.
As the specific memory layout is not important and potentially confusing for a classifier, we remove the addresses.
Memory operands are represented only by symbols, which we replace with their scrambled names.

Moreover, many operands represent memory locations indirectly, \eg{}, as offsets to registers.
When \texttt{objdump} recognizes a known address, it emits a comment containing the address and a symbol plus offset.
We replace the corresponding operand with the symbol and offset, also scrambled if applicable.
The \texttt{lea} instruction in \cref{lst:cwe191context} illustrates this replacement.

\subsubsection{Selecting and Representing Functions}
Every function in the \texttt{.text} section of an object file is a potential example for ROMEO, or part of the context of one (see \cref{sec:includingcontext}).
Hence, we extract a text representation of each function, consisting of a header line with the function's (scrambled) name, followed by the disassembly modified in the aforementioned manner.
The function name is prefixed with an exclamation mark, such that there is a unique token to mark the beginning of a function.
\Cref{lst:cwe191} shows an example of this representation.

However, not all functions are either positive or negative examples of a vulnerability.
There are also supporting or completely unrelated functions in the object files.
We only admit examples into the dataset that are primary or secondary good functions, or primary bad functions according to the regular expressions in \cite{JulietUserGuide}.
The remained is ignored.
We remove the primary good function to avoid label leakage (see the introduction to \cref{sec:extractingassembly}).
A large fraction, but not all, would be removed later in the process when duplicates are eliminated.

\subsubsection{Including Context}\label{sec:includingcontext}
The Juliet test suite lends itself to vulnerability classification on a function level, which we adapt for ROMEO.
However, individual functions cannot always be classified based on their body alone.
In terms of the source and sink model, a function containing a bad source might call another function containing the corresponding bad sink.
Its vulnerability status can then only be determined by analyzing the called function as well.

To mitigate this issue, we include context information by concatenating the text representation of a given function with the text representation of all functions that are referenced by it, recursively.
We exclude boilerplate and runtime functions such as \texttt{\_\_libc\_csu\_fini} as their content is always identical.
We further remove all bad functions if the given function is a primary or secondary good function, and vice versa, to avoid label leakage.

The scrambling of symbols applies to the context as well. \Cref{lst:cwe191} shows an example of a function and the relevant context, where it can be seen that the argument passed to it is actually used in the context of an I/O operation.

\subsection{Building a Dataset for Machine Learning}
The collection of examples from the previous section is then labeled either with binary (\enquote{good} or \enquote{bad}) or multi-class (CWE number or \enquote{no weakness}) labels as chosen by the user.
For our evaluation, we use the binary labels because most related work uses this formulation.
Subsequently, we eliminate duplicates and split the dataset for proper validation.

\subsubsection{Duplicate Elimination}\label{sec:duplicateelimination}
Duplicates in benchmarks based on the Juliet test suite are observed by several works \cite{allamanis2019adverse,Chakraborty2021DeepLearningBased,Russell2018AutomatedVulnerabilityDetection}.
Since each testcase in Juliet is initially roughly unique, duplicates likely result from preprocessing or extraction steps, or in our case, from the compilation process.
Our proposed ROMEO representation leads to a fraction of 2.6\% (with context) and 9.7\% (without context) duplicate examples, \ie{}, examples with an identical representation.

For each set of identical examples, we remove all but one instance.
Random selection determines which exact instance of a duplicate example is kept.

Our fraction of duplicates is lower compared to other works, \eg{}, \cite{Russell2018AutomatedVulnerabilityDetection}, which identifies an extreme value of 90.2\% for Juliet.
However, they do not remove the primary good functions of each testcase beforehand, which are mostly identical.
Moreover, their representation is designed to require a very small vocabulary, and removes most natural language elements such as identifiers and literals.
And most importantly, it does not consider context, which would help distinguish between similar instances.
An overview of duplicate fractions in vulnerability detection benchmarks can be found in \cite{Chakraborty2021DeepLearningBased}.

\subsubsection{Splitting}
We then randomly split the dataset into three parts for training, validation and testing, with a fraction of 80\%, 10\% and 10\% of the examples, respectively.
The resulting dataset is prepared for use by any machine learning methods capable of processing text sequences.
We name it ROMEO to highlight its dependence on Juliet.
It is publicly available at \url{https://gitlab.com/dlr-dw/romeo}.

\subsection{Descriptive Statistics of ROMEO}
This section is intended to give a brief overview of the distribution of samples in ROMEO.
We distinguish between ROMEO and ROMEO without context, since the duplicate elimination process (see \cref{sec:experiments}) affects each variant differently.
In \cref{tbl:examplesbyweakness}, we show the number of training, validation and test examples per weakness for the 20 most common weaknesses.
There are 91 CWEs in total.
However, the distribution of examples over CWEs is long-tailed, with CWE 190 accounting for approx. 11\% of all examples.
Flow variants are more uniformly distributed, hence the CWE distribution is more affected by the number of functional variants.
Overall, the ROMEO variant including context consists of 134129 training examples, 16766 validation examples and 16765 held-out test examples.
Without context, there are 124360, 15545 and 15544 examples, respectively.
In the training set with context, there 34831 positive and 99298 negative examples.
Without context, there are 34561 and 89799, respectively.

\addtolength{\tabcolsep}{-5pt}
\begin{table*}
  \centering{}
  \caption{Number of examples of top 20 weaknesses in ROMEO. Numbers differ depending on context inclusion because of subsequent duplicate elimination.}\label{tbl:examplesbyweakness}
  \begin{tabular}{llcccccc}
    & & \multicolumn{3}{c}{With Context} & \multicolumn{3}{c}{Without Context}\\
    \multicolumn{2}{l}{CWE} & Train & Val & Test & Train & Val & Test \\
    \hline{}%
    190 & Integer Overflow & 14794 & 1947 & 1878 & 14784 & 1863 & 1910 \\
    122 & Heap Based Buffer Overflow & 11480 & 1418 & 1395 & 9959 & 1235 & 1215 \\
    191 & Integer Underflow & 11149 & 1355 & 1375 & 11097 & 1374 & 1360 \\
    762 & Mismatched Mem. Mgmt. & 10533 & 1301 & 1332 & 10485 & 1238 & 1323 \\
    121 & Stack Based Buffer Overflow & 8828 & 1091 & 1084 & 7885 & 978 & 976 \\
    590 & Free Memory Not on Heap & 6166 & 750 & 754 & 4932 & 604 & 617 \\
    401 & Memory Leak & 4990 & 622 & 630 & 4965 & 633 & 603 \\
    134 & Uncontrolled Format String & 4482 & 551 & 573 & 4460 & 577 & 550 \\
    457 & Use of Uninitialized Variable & 3773 & 450 & 502 & 3780 & 476 & 466 \\
    124 & Buffer Underwrite & 3471 & 465 & 406 & 3032 & 387 & 365 \\
    127 & Buffer Underread & 3441 & 469 & 436 & 3016 & 397 & 369 \\
    369 & Divide by Zero & 3299 & 399 & 419 & 3255 & 413 & 427 \\
    195 & Signed-Unsigned Conv. Error & 3211 & 384 & 399 & 2586 & 317 & 319 \\
    194 & Unexpected Sign Extension & 3164 & 410 & 420 & 2599 & 312 & 305 \\
    415 & Double Free & 2904 & 345 & 380 & 2929 & 319 & 369 \\
    400 & Resource Exhaustion & 2811 & 312 & 342 & 2766 & 347 & 345 \\
    126 & Buffer Overread & 2728 & 335 & 330 & 2471 & 290 & 287 \\
    36 & Absolute Path Traversal & 2668 & 341 & 330 & 2327 & 303 & 296 \\
    23 & Relative Path Traversal & 2651 & 326 & 361 & 2524 & 304 & 341 \\
    78 & OS Command Injection & 2647 & 332 & 313 & 2469 & 312 & 281\\
    \hline{}%
    & Total & 134129 & 16766 & 16765 & 124360 & 15545 & 15544 \\
  \end{tabular}
\end{table*}
\addtolength{\tabcolsep}{5pt}

\begin{lstfloat*}
\centering{}
\begin{subfigure}[]{0.45\textwidth{}}
\begin{lstlisting}
!lc383:
push rbp
mov rbp,rsp
sub rsp,0x10
mov DWORD PTR [rbp-0x4],0x0
mov DWORD PTR [rbp-0x4],0x0
mov eax,DWORD PTR [rbp-0x4]
sub eax,0x1
mov DWORD PTR [rbp-0x8],eax
mov eax,DWORD PTR [rbp-0x8]
mov edi,eax
call lc188
leave
ret
\end{lstlisting}
\caption{The extracted function.}
\end{subfigure}%
\begin{subfigure}[]{0.45\textwidth{}}
\begin{lstlisting}
!lc188:
push rbp
mov rbp,rsp
sub rsp,0x10
mov DWORD PTR [rbp-0x4],edi
mov eax,DWORD PTR [rbp-0x4]
mov esi,eax
lea rdi,_IO_stdin_used+0x6e
mov eax,0x0
call printf
nop
leave
ret
\end{lstlisting}
\caption{The context of the extracted function.}\label{lst:cwe191context}
\end{subfigure}
\caption{A function extracted from a Juliet testcase concerning CWE-191 (Integer Underflow) and its accompanying context. Both are in the text representation as described in \cref{sec:extractingexamples}. This example illustrates one purpose of including context functions, namely to check whether a result is actually used, \eg{}, in an API call.}\label{lst:cwe191}
\end{lstfloat*}



\section{Experiments}\label{sec:experiments}
We consider the ROMEO dataset and representation itself our main contribution for this work.
Still, there are research questions relating to the design goals of ROMEO and the usefulness of its representation for vulnerability detection applications that should be answered empirically.
We identify the following research questions:
\begin{description}
  \item[RQ1] What benefits and drawbacks are associated with the inclusion of context (as described in \cref{sec:includingcontext})?
  \item[RQ2] To what extent, if any, does our assembly language representation leak label information?
  \item[RQ3] How does an off-the-shelf model using our assembly language representation compare to other methods, including ones with access to the C/C++ source code?
\end{description}

In the remainder of this section, we first describe our experimental setup including datasets and baselines.
We then present our results structured along the aforementioned research questions and offer brief discussions.
Finally, we summarize the results, linking questions and answers, and address the limitations of this evaluation.

\subsection{Setup}
This section describes the situation concerning binary vulnerability detection datasets derived from Juliet as well as our Transformer-based vulnerability detector.

\subsubsection{Datasets}\label{sec:otherdatasets}
In \cref{sec:relatedwork}, we list several works that perform vulnerability detection on assembly or machine code representations of the Juliet test suite.
In particular, there is \cite{Le2019MaximalDivergenceSequential}, which is only available in an already encoded vector form, from which the original instructions cannot be recreated.
The same is true for \cite{Lee2017LearningBinaryCode}.
BVATT \cite{Afanador2021BenchmarkFrameworkSupport,Afanador2020RepresentativenessBenchmarkVulnerability}, is not available publicly anymore, against the claims in their work.

In all cases, the respective authors did not accommodate our request for the datasets in their original form.

\subsubsection{Baselines}\label{sec:baselines}
While we cannot directly compare our representation combined with the Transformer based approach (see \cref{sec:transformers}) to other methods on identical data (due to availability reasons explained in \cref{sec:otherdatasets}), we can draw conclusions from comparisons to other work on reasonably similar data.
Specifically, we compare our approach to works that evaluate their methods on test sets derived from the Juliet test suite, namely Instruction2Vec \cite{Lee2017LearningBinaryCode,Lee2019Instruction2vecEfficientPreprocessor}, BVDetector \cite{Tian2020BVDetectorProgramSlice}, Russel \etal{} \cite{Russell2018AutomatedVulnerabilityDetection}, and \textsc{ReVeal} \cite{Chakraborty2021DeepLearningBased}. We modify the ROMEO dataset to match the respective subset or construction as closely as possible in each case.

\subsubsection{Transformer-based Model for Vulnerability Detection}\label{sec:transformers}
The empirical part of this work focuses mainly on exploring the advantages and drawbacks of an assembly language representation for vulnerability detection.
Hence, we select a rather generic, but very powerful approach to perform the actual classification task, namely Transformers~\cite{Vaswani2017AttentionIsAll}.
Specifically, we use the pre-trained CodeBERT model~\cite{Feng2020CodeBERTPretrained} for initialization.
Our implementation is based on PyTorch~\cite{Paszke2019PyTorchImperativeStyle} and HuggingFace Transformers~\cite{Wolf2020HuggingFaceTransformers}.

Because assembly language is less complex and variable than the languages in the initial training set of CodeBERT (which does not include assembly directly), the included tokenization and encoding is not optimal.
Hence, we replace them with a byte-pair encoding~\cite{Devlin2018BERTPretraining} optimized on the ROMEO training set, which can represent common mnemonics such as \texttt{mov} by a single token.
With context, the average example is 318.4 tokens long, where the model can handle at most 512 tokens. Without context, it is 157.2 tokens.

In our comparison with other methods, we use the name \enquote{ROMEO method} to refer to the combination of our ROMEO assembly language representation and the Transformer model.

We train the model for ten epochs on the ROMEO training set with and without context, respectively.
One epoch equals one iteration over the entire training set.
We use a minibatch size of 16, a learning rate of 1.1e-5, and a L2 regularization coefficient of 3e-4.
These values are determined using the validation set, and subsequent evaluations are performed on the held-out test set.
Each training and evaluation is performed three times, and all reported results are the average over all runs, with standard deviations displayed where applicable.

\subsection{Results}
In the following, we present our experimental results. Each section addresses one of the three research questions from the beginning of \cref{sec:experiments}, in order.
All metrics are reported in terms of a binary classification problem (vulnerable / not vulnerable) for better comparison with other methods.
However, the ROMEO dataset can easily be configured by users to include multi-class classification labels, as can the ROMEO method.

\subsubsection{Context}
\begin{table*}
  \centering{}
  \caption{Accuracy of our Transformer-based method on the held-out test set of ROMEO. We list the five CWEs where including context has the strongest positive or negative effect on overall accuracy, respectively. The number of examples is provided to assess the overall effect size.}\label{tbl:contextaccuracybyweakness}
  \begin{tabular}{llcccc}
    & & \multicolumn{2}{c}{Accuracy (\%)} & \multicolumn{2}{c}{Examples (\#)}\\
    \multicolumn{2}{l}{CWE / Context:}  & w & w/o & w & w/o \\
    \hline{}%
    190 & Integer Overflow & 98.3 & 89.2 & 1878 & 1910 \\
    191 & Integer Underflow & 98.1 & 88.7 & 1374 & 1360 \\
    762 & Mismatched Mem. Mgmt. & 97.9 & 89.9 & 1332 & 1326 \\
    122 & Heap Based Buffer Overflow & 95.7 & 88.2 & 1395 & 1215 \\
    401 & Memory Leak & 96.8 & 89.5 & 631 & 604 \\
    \hline{}%
    676 & Use of Potentially Dangerous Func. & 88.9 & 95.2 & 6 & 5 \\
    468 & Incorrect Pointer Scaling & 75.0 & 82.4 & 8 & 17 \\
    396 & Catch Generic Exception & 73.0 & 77.2 & 18 & 19 \\
    590 & Free Memory Not on Heap & 99.7 & 99.8 & 753 & 619 \\
    398 & Poor Code Quality & 94.3 & 96.6 & 64 & 59 \\
    \hline{}%
    & Overall/Total  & 96.9 & 90.2 & 16764 & 15544 \\
  \end{tabular}
\end{table*}

\addtolength{\tabcolsep}{-4pt}

\begin{table*}
  \centering{}
  \caption{Accuracy of our Transformer-based method on the held-out test set of ROMEO. We list the five flow variants, where including context has the strongest positive or negative effect on overall accuracy, respectively. The number of examples is provided to assess the overall effect size.}\label{tbl:contextaccuracybyflowvariant}
  \begin{tabular}{lp{0.65\textwidth}cccc}
    & & \multicolumn{2}{c}{Acc. (\%)} & \multicolumn{2}{c}{Ex. (\#)}\\
    \multicolumn{2}{l}{Flow Variant / Context:}  & w & w/o & w & w/o \\
    \hline{}%
    62 & Data flows using a C++ reference from one function to another in different source files & 99.2 & 66.2 & 237 & 203 \\
    42 & Data returned from one function to another in the same source file & 98.6 & 65.7 & 236 & 188 \\
    61 & Data returned from one function to another in different source files & 99.0 & 70.0 & 242 & 217 \\
    43 & Data flows using a C++ reference from one function to another in the same source file & 97.7 & 65.9 & 219 & 204 \\
    83 & Data passed to a class constructor and destructor by declaring the class object on the stack & 92.6 & 72.3 & 332 & 325 \\
    \hline{}%
    8 & if(staticReturnsTrue()) and if(staticReturnsFalse()) & 98.7 & 98.4 & 501 & 516 \\
    15 & switch(6) and switch(7) & 99.3 & 99.0 & 451 & 442 \\
    14 & if(globalFive==5) and if(globalFive!=5) & 99.3 & 99.2 & 511 & 504 \\
    5 & if(staticTrue) and if(staticFalse) & 99.1 & 99.1 & 501 & 493 \\
    21 & Flow controlled by value of a static global variable. All functions contained in one file. & 99.3 & 99.8 & 304 & 317 \\
    \hline{}%
    & Overall/Total  & 96.9 & 90.2 & 16764 & 15544 \\
  \end{tabular}
\end{table*}

\addtolength{\tabcolsep}{4pt}

RQ1 asks \enquote{what benefits and drawbacks are associated with the inclusion of context?}
We answer this research question quantitatively and qualitatively using two variants of the ROMEO dataset, one with context and one without.
Context is defined in \cref{sec:includingcontext}.

We first apply our ROMEO method (see \cref{sec:transformers}) to both variants of the ROMEO dataset.
With context, the overall accuracy on the held-out test set is 96.9\% and the overall F1 score is 94.0\%.
Without context, the accuracy and F1 score are 90.2\% and 81.9\%, respectively.
Hence, on average, our method strongly benefits from the included context.

From a machine learning point of view, removing the context information can be interpreted as manual feature selection, which is occasionally done on purpose to prevent overfitting.
Removing features that are unrelated to the problem prevents the classifier from adapting to spurious correlations.
However, in our case, there are clearly instances where context is crucial to determine whether a function is vulnerable or not, \eg{}, when the actual vulnerability is spread over multiple function calls.
Still, context information might not always be relevant and introduce misleading features.
To get a clearer picture of this situation, we evaluate our method for each flow variant and CWE individually.

\Cref{tbl:contextaccuracybyflowvariant} shows the accuracy obtained with and without context information and which flow variants are most affected.
In line with our expectations, the flow variants that benefit most from including context all describe vulnerabilities spread over multiple functions or even multiple translation units.
Here, we see accuracy improvements up to 34.5 percent points.
Without context, it is not possible to identify whether a function is calling a \enquote{bad sink} (assuming no label leakage through symbols or inlining by the compiler).
We also show the flow variants that are most negatively affected, and in the worst case, the difference in accuracy is less than two percent points.

In \cref{tbl:contextaccuracybyweakness}, we present the CWE types most affected by the inclusion of context.
Overall, the most positively affected CWE types are simply those with the most examples and vice versa for the most negatively affected types.
The most negatively affected weakness types have such low representation in Juliet that we cannot extract a meaningful interpretation of the results.
In contrast to the flow variants, there is no set of CWE types that specifically benefit from context inclusion -- most of them do.

To answer RQ1, including context in the assembly language representation of ROMEO is clearly beneficial for vulnerability detection.
The benefits and drawbacks of including context do not appear to be specific to certain types of weaknesses in terms of CWEs.
They are specific to certain presentations of vulnerabilities, \ie{}, flow variants, that are spread out over multiple functions or translation units.

\subsubsection{Label Leakage}
RQ2 asks \enquote{To what extent, if any, does our assembly language representation leak label information?}
Preventing label leakage is also one of the design goals of ROMEO (see \cref{sec:extractingassembly}).

Our process is carefully designed to prevent label leakage in terms of features, \eg{} by renaming symbols (\cref{sec:symbolrepresentation}) and excluding telltale functions from the context (\cref{sec:includingcontext}).
Still, it is theoretically possible that some occurrence of label leakage is missing.
However, we can estimate the label leakage in ROMEO empirically.
The Juliet test suite includes examples of CWE-546 (suspicious comment), whose vulnerability status is impossible to infer in our assembly language representation, because the comments from C/C++ are not carried over.
In the test set, there are six positive and 25 negative examples of this CWE. If there is no label leakage, the highest possible accuracy any method can achieve for this CWE is 80.6\%, by constantly making negative predictions.
Our method reaches 96.9\% accuracy on the whole dataset, but achieves only 77.4\% on CWE-546, which is below this threshold and a strong indication that there is no exploited label leakage.


However, label leakage can also be present on a dataset level even if the individual examples are free of it.
While ROMEO does not have a time component, there are three distinct groupings that should be considered during the sampling process, namely CWEs, flow variants, and functional variants.
For our evaluation, we require all CWEs and flow variants to be represented proportionally in all splits, which precludes splitting along them.
Testing generalization across CWEs or flow variants could be an interesting investigation, but is beyond the scope of this work.
Furthermore, we do not split the dataset along the functional variants because it would hinder our comparison with other work (see \cref{sec:comparisonwithothermethods}), all of which apply random sampling.

To summarize: empirically, our representation does not leak label information.

\subsubsection{Comparison with Other Methods}\label{sec:comparisonwithothermethods}
\begin{table*}
  \centering{}
  \caption{Accuracy and F1 score on the held-out test set of ROMEO with and without context, compared to other methods on their respective variants of Juliet.
  \textit{Note that Russel \etal{} works on slices, while ROMEO and }\textsc{ReVeal}\textit{ work on functions.}}\label{tbl:comparisonrussell}
  \begin{tabular}{llcc}
    Method & Dataset & Accuracy (\%) & F1 (\%)\\
    \hline{}%
    ROMEO & ROMEO w/o context & 90.2 $\pm$ 0.2 & 81.9 $\pm$ 0.4 \\
    ROMEO & ROMEO & 96.9 $\pm$ 0.2 & 94.0 $\pm$ 0.4\\
    Russell \etal{} & Juliet (slices) & --- & 84.0\\
    \textsc{ReVeal} & Juliet (functions, no SMOTE) & --- & 93.7
  \end{tabular}
\end{table*}

\begin{table*}
  \centering{}
  \caption{Accuracy and F1 score on subsets of the held-out test set of ROMEO, compared to BVDetector on similar subsets of the Juliet test suite.
  \textit{Note that BVDetector works on slices, while ROMEO works on functions.}}\label{tbl:comparisonbvdetector}
  \begin{tabular}{llcc}
    Method & Dataset & Accuracy (\%) & F1 (\%)\\
    \hline{}%
    ROMEO & ROMEO (MC) & \textbf{95.6} $\pm$ 0.5 & \textbf{91.3} $\pm$ 1.1\\
    BVDetector & Juliet (MC, slices) & 94.8 & 85.4 \\
    \hline{}%
    ROMEO & ROMEO (NH) & \textbf{98.1} $\pm$ 0.1 & \textbf{96.1} $\pm$ 0.2 \\
    BVDetector & Juliet (NH, slices) &  97.6 & 92.2 \\
    \hline{}%
    ROMEO & ROMEO (MC+NH) & \textbf{97.1} $\pm$ 0.2 & \textbf{94.1} $\pm$ 0.5\\
    BVDetector & Juliet (MC+NH, slices) & 96.7 & 89.9
   \end{tabular}
\end{table*}

RQ3 asks \enquote{How does an off-the-shelf model using our assembly language representation compare to other methods, including ones with access to the C/C++ source code?}

As alluded to in \cref{sec:baselines}, we can only offer an estimate of relative performance.
All performance numbers that we use from other works are based on the same Juliet test suite, but there are different evaluation processes.
The works differ in granularity, \ie{}, program slices \cite{Tian2020BVDetectorProgramSlice} vs. functions \cite{Lee2017LearningBinaryCode,Russell2018AutomatedVulnerabilityDetection,Chakraborty2021DeepLearningBased}.
Moreover, we extract sample functions from the full dataset unconditionally, but popular options include a selection of candidates by static analysis.
Deduplication is performed at different stages or according to different representations \cite{Russell2018AutomatedVulnerabilityDetection,Chakraborty2021DeepLearningBased}, or not at all \cite{Lee2017LearningBinaryCode,Tian2020BVDetectorProgramSlice}.
With these limitations in mind, we offer comparisons to related methods \cite{Lee2017LearningBinaryCode,Russell2018AutomatedVulnerabilityDetection,Tian2020BVDetectorProgramSlice,Chakraborty2021DeepLearningBased} and modify our dataset to match their evaluations as closely as possible.

\paragraph{Instruction2Vec + Text-CNN} This combination of methods \cite{Lee2017LearningBinaryCode,Lee2019Instruction2vecEfficientPreprocessor} is evaluated on a small subset of the Juliet test suite, namely vulnerabilities of type CWE-121 (stack-based buffer overflows).
It also relies on assembly language instructions instead of source code.
For comparison, we evaluate our Transformer-based method on this single vulnerability category only.
Our ROMEO method achieves an accuracy of 97.3\%, compared to 96.1\% reported by Lee \etal{}
Both methods operate on a function level.

\paragraph{Russel \etal{}} \cite{Russell2018AutomatedVulnerabilityDetection} employs a bespoke C/C++ lexer to obtain a token sequence directly from the source code.
On the Juliet suite, which is not subsampled in their work, they report an F1 score of 84.0\% for the best combination of classifiers.
In contrast, the ROMEO method obtains an F1 score of 94.0\% over the whole suite (see also \cref{tbl:comparisonrussell}).

\paragraph{BVDetector} \cite{Tian2020BVDetectorProgramSlice} extracts program slices instead of functions, but operates on assembly language instructions.
The authors evaluate their method on a subset of Juliet involving memory corruption (MC) and number handling (NH) vulnerability types as defined by the STONESOUP program\footnote{\url{https://samate.nist.gov/SARD/around.php}}.
We evaluate our ROMEO method on a subset corresponding to these types and obtain an accuracy of 97.1\% on MC and NH combined, where BVDetector reaches 96.7\%.
Both subsets are evaluated individually in \cref{tbl:comparisonbvdetector}.

\paragraph{\textsc{ReVeal}} \cite{Chakraborty2021DeepLearningBased} is proposed alongside an in-depth analysis of the current state of deep learning-based vulnerability detection and an eponymous dataset.
It provides interesting insight into the limitations of synthetic datasets such as Juliet, but nevertheless the authors evaluate their method on it.
On the unmodified dataset, they report an F1 score of 93.7\%, compared to ROMEO's very close 94.0\% (see also \cref{tbl:comparisonrussell}).
However, due to the extreme class imbalance observed by them, they apply SMOTE to alter the distribution by resampling that data and increase their F1 score to 95.7\%.

\paragraph{}
Overall, the ROMEO method, consisting of our assembly language representation combined with an off-the-shelf Transformer approach, performs better than other state-of-the-art methods, except for \textsc{ReVeal}-SMOTE, to the degree that these methods can be compared fairly.
Importantly, the outperformed methods include \cite{Russell2018AutomatedVulnerabilityDetection,Chakraborty2021DeepLearningBased}, which have access to the full C/C++ source code including comments.

\subsection{Limitations of this Study}\label{sec:limitations}
There are aspects in which this investigation is limited.
For example, our evaluation is performed on the function level, similar to \cite{Lee2017LearningBinaryCode,Russell2018AutomatedVulnerabilityDetection,Chakraborty2021DeepLearningBased}. This design choice is appropriate for the Juliet test suite and our derived dataset ROMEO, but may not translate well to real-world applications.
This limitation applies only to our dataset, since our representation could in principle be annotated on a per-instruction level. Alternatively, explainable methods such as IVDetect~\cite{Li2021VulnerabilityDetectionFine} could be applied to generate per-line results even from per-method annotations.

We maintain that the Juliet test suite is a sensible choice for our investigation as it allows for a methodical and detailed evaluation of our representation. It also covers a very wide range of weaknesses. However, it is not representative of real-world software projects in terms of vulnerability statistics and code complexity.

Moreover, in its present state, the examples in the dataset only include information from \texttt{.text} sections. Hence, information from data segments is missing. In \cref{lst:cwe191}, this can be seen in the context function \texttt{lc188}. The format string handed to \texttt{printf} is not part of the representation.

Currently, comments from the C/C++ code are also not carried over to the assembly language representation. While this could be solved by accessing debug information, it is not reasonable to expect this information to be present in a real-world binary analysis scenario.
Our choice of x86-64 ISA is another minor limitation, however, our representation is not technically restricted to it and could be adapted, \eg{}, to ARM.

We discuss proposed solutions to these limitations in \cref{sec:futurework}.

\subsection{Summary of Results}
In the following, we provide a brief summary of the research questions posed and answered in this evaluation:
\begin{description}
  \item[RQ1] What benefits and drawbacks are associated with the inclusion of context (as described in \cref{sec:includingcontext})?
  \item[RA1] Including context in the assembly language representation of ROMEO is clearly beneficial for vulnerability detection.
  The benefits are specific to certain presentations of vulnerabilities that are spread out over multiple functions.
  \item[RQ2] To what extent, if any, does our assembly language representation leak label information?
  \item[RA2] Empirically, our representation does not leak label information. 
  \item[RQ3] How does an off-the-shelf model using our assembly language representation compare to other methods, including ones with access to the C/C++ source code?
  \item[RA3] It performs better than other state-of-the-art methods (including C/C++-based), except for \textsc{ReVeal}-SMOTE, to the degree that these methods can be compared fairly. 
\end{description}
\section{Conclusion}\label{sec:conclusion}
In this work, we present ROMEO, a publicly available, reproducible and reusable binary vulnerability detection benchmark dataset.
It is derived from the Juliet test suite \cite{JulietUserGuide} of C/C++ vulnerabilities by compiling and extracting an assembly language representation.

Our representation incorporates context information in the form of related functions, which allows for detection of cross-function vulnerabilities.
It also includes symbols to relate API call semantics.
However, symbols that would leak label information are replaced with random names.

To evaluate the capabilities of this representation, we combine it with an off-the-shelf Transformer model to build the ROMEO method, and compare the combination with other methods.
We show that even state-of-the-art methods that have full access to the C/C++ code only outperform ROMEO in specific cases.
Furthermore, we find that, empirically, there is no label leakage in the ROMEO dataset or caused by our representation.
Finally, there is a clear benefit to including context functions derived from the call graph for types of vulnerabilities that are span multiple functions.

In the following, we provide a brief outlook regarding possible future research.

\subsection{Future Work}\label{sec:futurework}
First and foremost, future work should address the limitations mentioned in \cref{sec:limitations}. Low-hanging fruit includes adding data to the context instead of only code. Currently, only the \texttt{.text} segment is analyzed. This analysis could be extended to \texttt{.data}.

Insight could be gained from adding source code comments to the appropriate location in the assembly representation and adding further details using debug information.
However, this needs to be done carefully as to not introduce label leakage.
A further interesting evaluation would involve repeating the experiments on a different ISA, \eg{}, ARM, and observing the qualitative effects.

On a larger scale, our process of deriving a dataset from binaries could be adapted to real-world vulnerabilities, \eg{} from the NVD.
However, it is not trivial to construct a dataset of C/C++ code vulnerabilities that can be fully compiled, which is a requirement for binary vulnerability detection. Alternatively, one could adapt an existing dataset, \eg{}, Big-Vul~\cite{Fan2020C/CCodeVulnerability}, and accept that only a subset is provided in a readily compilable form.



\section*{Availability}
We provide the entire dataset, including the raw binaries and source code to generate them and the representation, and to reproduce all results in this work.
The dataset is available for download at \url{https://gitlab.com/dlr-dw/romeo}.


\bibliographystyle{elsarticle-harv}
\bibliography{paper}

\end{document}